# Effect of heterostructure engineering on electronic structure and transport properties of two-dimensional halide perovskites

Rahul Singh,[1] Prashant Singh[2,*] and Ganesh Balasubramanian[3]

[1]Department of Mechanical Engineering, Iowa State University, Ames, IA 50011
[2]Ames Laboratory, U.S. Department of Energy, Iowa State University, Ames, IA 50011
[3]Department of Mechanical Engineering & Mechanics, Lehigh University, Bethlehem, PA 18015

**Abstract**

Organic-inorganic halide perovskite solar cells have attracted much attention due to their low-cost fabrication, flexibility, and high-power conversion efficiency. More recent efforts show that the reduction from three- to two-dimensions (2D) of organic-inorganic halide perovskites promises an exciting opportunity to tune their electronic properties. Here, we explore the effect of reduced dimensionality and heterostructure engineering on the intrinsic material properties, such as energy stability, bandgap and transport properties of 2D hybrid organic-inorganic halide perovskites using first-principles density functional theory. We show that the energetic stability is significantly enhanced by engineered perovskite heterostructures that also possess excellent transport properties similar to their bulk counterparts. These layered chemistries also demonstrate the advantage of a broad range of tunable bandgaps and high-absorption coefficient in the visible spectrum. The proposed 2D heterostructured material holds potential for nano-optoelectronic devices as well as for effective photovoltaics.

**Corresponding author**: psingh84@ameslab.gov

**Introduction**

Over the last decade, perovskites have been one of the most intensely examined classes of materials because of their outstanding optoelectronic properties [1-6]. The versatility of these materials encompasses a series of optoelectronic devices such as light-emitting diodes [1,2], transistors [3], lasing applications [4], as well as other intriguing electronic properties [5-11]. In particular, organic-inorganic halide perovskites (OIHP), archetypically $CH_3NH_3PbX_3$ (X = Cl, Br or I), have attracted significant attention because of their remarkable photovoltaic properties [12-14], achieving power conversion efficiency as high as 22.1% [4]. Thin OIHP films that are typically synthesized and examined for the electronic transport, suffer from stability issues. A potential solution is to employ planar heterostructures in lieu of the 3-dimensional (3D i.e., bulk) films [13,15-23]. Quantum confinement effect in two-dimensional (2D) chemistries [24] increases the bandgap due to a blue shift that decreases with increasing number of layers in a heterostructured material [25-28]. The decrease in bandgap due to increasing number of inorganic layers is detrimental to the stability of the heterostructure.

Recently, there has been a renewed interest in thermoelectric properties of halide-perovskites, which was motivated by the realization that complexity at multiple length scales can lead to new transport mechanisms in high performance perovskite materials. Theoretical predictions suggest that the thermoelectric efficiency could be greatly enhanced by quantum confinement [29] due to heterostructure engineering [30,31]. The formation of heterostructures by chemical layer deposition has been attempted to enhance photoluminescence [32], charge-transfer mechanism [33], and surface dopants [33]. The chemical layer deposition can lead to quantum confinement (QC) of charges. The decreasing dimensionality with QC aids in narrowing down electron energy bands, which produces high effective masses and Seebeck coefficients. Moreover, similar sized heterostructures decouple the Seebeck coefficient and electrical conductivity by electron filtering [34] that could result in an improved thermoelectric energy conversion.

In this work, we employed first-principles density functional theory (DFT) to examine the stability, electronic-structure, transport and optical properties of [001] terminated 2D OIHPs [35]. The 2D assemblies of $MAPbI_3$ (MA = $CH_3NH_3$) were created from the bulk

OIHPs and stacked on monolayer (ML) MoS$_2$ (MoS$_2$ML). The thermodynamic stability of these heterostructure assemblies was analyzed with respect to MAPbI$_3$ layer distance from monolayer MoS$_2$ (MoS$_2$ML). This study reveals that heterostructure of 2D-MAPbI$_3$ engineered with 2D-MoS$_2$ remarkably improves the thermodynamic stability of OIHPs. A detailed electronic-structure, bandgap, and transport properties study were performed on thermodynamically stable MAPbI$_3$ML/MoS$_2$ML hetero-assembly. A direct bandgap within the optimal range does not guarantee desired absorption in the visible range, therefore, understanding optical absorption also becomes critical in solar-cell materials for future applications.

**Computational Methods**

**Supercell generation.** We use relaxed 2D-MoS$_2$ unit cell (3 atom/cell) and relaxed 2D-MAPbI$_3$ (15 atom/cell) to construct a common supercell with 174 atoms (6 C; 6 N; 36 H; 12 Pb; 30 I; 28 Mo and 56 S atoms in the supercell). We generate a common cell in such a way that we can achieve minimum strain, less than 1%.

**DFT Method.** We examine bulk and 2D variants of MA(Pb/Sn)I$_3$ OIHPs using first-principles density functional theory (DFT) [36,37] using the Vienna *ab initio* simulation package (VASP) [38-40]. We construct [001] terminated 2D OIHPs from 3D MA(Pb/Sn)I$_3$. For geometry optimization and electronic structure calculations, we use the projected augmented-wave (PAW) basis [38] and the Perdew–Burke–Ernzerhof (PBE) [40] exchange-correlation functional. We also use self-consistent GW potential functional of higher accuracy to establish the robustness of our predictions. The charge and forces are converged to $10^{-5}$ eV and 0.01eV/Å, respectively, using energy cut-off of 800 eV. The Monkhorst-Pack [41] k-mesh grid of 7×7×3 is used for 2D MAPbI$_3$, 2D MASnI$_3$, and 3×3×5 for MAPbI$_3$ML/MoS$_2$ML. The tetrahedron method with Blöchl corrections is used to calculate the density of states (DOS) [38]. The thermoelectric properties are calculated using the BoltzTrap [42] code interfaced with VASP. The absorption spectra are calculated by obtaining the frequency dependent dielectric matrix using VASP [43].

**Results and Discussion**

**Cohesive and formation energies.** We calculate the cohesive energies using the van der Waals (vdW) density functional (vdW-DF), which correctly describes the vdW interaction in molecular complexes or solids at a reasonable computational cost [44,45]. We use a modified vdW-DF with the recently proposed rev-vdWDF2 [45]. The cohesive energy per atom was calculated as

$$E_c = E_{total} - \sum E_i^{isolated}$$

where energy ($E_{total}$) is calculated for the optimized heterostructures, $E_c$ is the cohesive energy of the heterostructures with respect to constituent elements, where $E_i^{isolated}$ are the atomic energies of isolated with 'i' being the atom index. The cohesive energies were estimated both with and without vdW (dispersive potential) as shown in Table 1. Lower energy per atom in Table 1 with vdW shows that it has significant effect on heterostructure energies.

**Table 1**. Cohesive energies with and without vdW interaction for MAPbI$_3$ML/MoS$_2$ML.

| System/E (eV/atom) | $E_c$ | $E_c$ (vdW) |
|---|---|---|
| **MoS$_2$ML/MAPbI$_3$ML** | -1.66E+00 | -4.60E+00 |

The formation energy ($E_{form}$) for the different heterostructures was calculated as

$$E_{form} = E_{total} - \sum E_i$$

where $E_i$ is the ground state energy per atom (elemental solid) of the constituent elements in bulk form, where i = C (hexagonal), H (hexagonal), Pb (face-centere cubic), I (orthorhombic), S (orthorhombic) and Mo (body-centered cubic). The $E_{form}$ of MAPbI$_3$ML/MoS$_2$ML at different interlayer distances is shown in Table 2. Evidently, MAPbI$_3$ML/MoS$_2$ML at z (=3.22 Å) is energetically most stable and hence we focus on this case throughout this work.

**Table 2**. Formation energy (meV/atom) with varying MAPbI$_3$ layer distance from the MoS$_2$-layer, where z=3.22 Å.

| MoS$_2$-ML/MAPbI$_3$/z | z - 0.5 | z | z + 0.5 |
|---|---|---|---|
| E$_f$ (meV/atom) | +39.100 | -14.700 | -0.393 |

*Structural analysis of bulk and 2D variants of MAYI$_3$ (Y=Pb, Sn):*

(a)

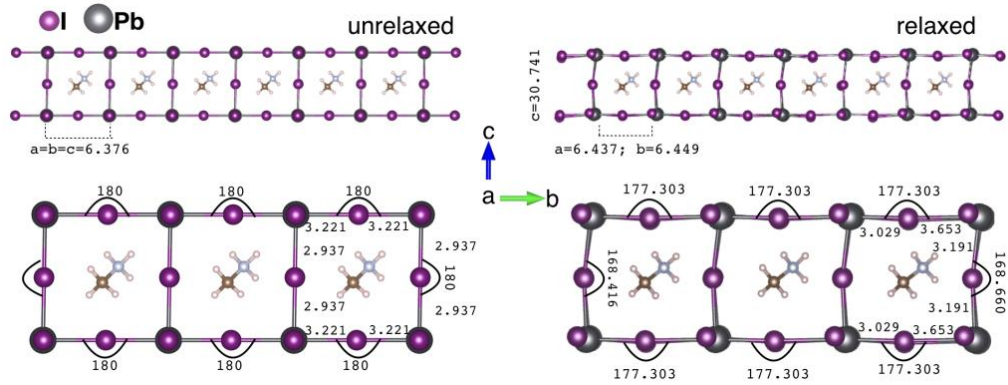

(b)

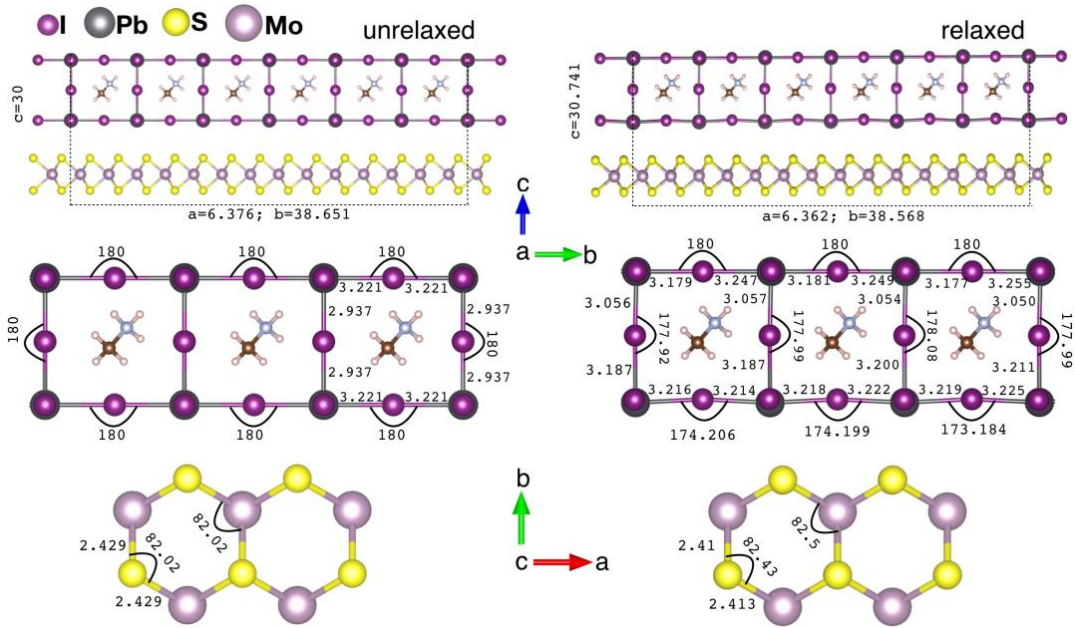

**Figure 1.** The unrelaxed (left-panel) and optimized structure (right-panel) of (a) MAPbI$_3$ monolayer; and (b) MAPbI$_3$ML/MoS$_2$ML. The average bond-length and bond-angles of MoS$_2$ML changes from 2.428 Å (unrelaxed) to 2.412 Å (relaxed) and 82.02° (unrelaxed) to 82.5° (Mo-S-Mo)/82.43° (S-Mo-S), respectively.

Hybrid organic-inorganic perovskites have an ABX$_3$ architecture, where A is a monovalent organic cation, CH$_3$NH$_3^+$ (i.e., MA$^+$), while B is a metal cation (i.e., Pb$^{2+}$, Sn$^{2+}$), and X is a halide anion (i.e., Cl, Br, I or their mixtures). In typical perovskite crystals, B occupies the center of an octahedral [BX$_6$] cluster [47], while A is 12-fold cuboctahedrally coordinated with X anions [15,48]. Generally, A does not directly play a major role in determining the band-structure, but its size is important. We construct a 2D structure (Figure 1) by cutting through PbI$_2$ plane [001] of bulk MAPbI$_3$ and expose the PbI$_2$-terminated surface. We maintain the thicknesses of the slab to one monolayer along [001] and study the effect of heterostructure on the stability and electronic properties.

Here, we adopt the high temperature pseudo-cubic phase of MAPbI$_3$ [49,50]. The calculated (experimental) [51,52] lattice constants of bulk MAPbI$_3$ are $a$ = 6.432 (6.361) Å, $b$ = 6.516 (6.361) Å, $c$ = 6.446 (6.361) Å, and α ≈ β ≈ γ ≈ 90°. The predictions indicate a 3.41% increase in the simulated equilibrium cell volume relative to that in the experiments [51,52] with 1.27% increase in average Pb-I bond length. For 2D MAPbI$_3$, the calculated (experimental) lattice parameters of the PbI$_2$ surface are $a$ = 6.437 (6.361) Å and $b$ = 6.449 (6.361) Å respectively. We find a 1.38% increase in the lattice constant in [110] terminated surface of the simulated material, although the thickness of the slab shrinks by 2.5%. The reduced thickness (Fig. S1 (c), Fig. 1(a)) of the slab shrinks the overall volume by 1.33% relative to the bulk MAPbI$_3$ in the experiments.

The structural parameters of optimized MoS$_2$-deposited MAPbI$_3$ML are presented in Figure 1. The MAPbI$_3$/MoS$_2$ monolayer supercell consists of 174 atoms/per unit cell (C=6; N=6; H=36; Pb=12; I=30; Mo=28; S=56 atoms). The unit cell is periodic in the *x-y* plane and vacuum of at least 18 Å is maintained in the *z*-direction. The unit cell is constructed such that there is less than 1% mismatch between MoS$_2$ (2 atom per cell) and MAPbI$_3$ (12 atom per cell). We further relax the supercell to remove any residual strain from the unit cell construction due to lattice mismatch. The (X×Y×S) dimensions of the cell after relaxation are (6.362 Å × 38.567 Å × 30.741 Å). The optimized lattice constants *a* and *b* are 6.362 (6.361) Å and 6.428 (6.361) Å [53,54], respectively, similar to that of bulk MAPbI$_3$, shown in Figs. 1. Likewise, the average Pb-I bond length is 3.201 Å is similar to that in the bulk MAPbI$_3$. The distribution of Pb-I bond length tends to be uniform only for the bulk and the MoS$_2$ML deposited MAPbI$_3$. However, the change of Pb-I-Pb/I-Pb-I bond-angles in 2D and MAPbI$_3$ML/MoS$_2$ML are comparable with those of bulk MAPbI$_3$. To note, the bond lengths of Pb-I are substantially enhanced from bulk to 2D and longer than experimental (bulk) MAPbI$_3$. We notice 1.2% per formula unit volume decrease in optimized MAPbI$_3$ML/MoS$_2$ML compared to 2D MAPbI$_3$. The reduced volume leads to stronger intralayer hybridization. At the MoS$_2$ML/MAPbI$_3$ML interface, MAPbI$_3$ shows negligible polyhedral distortion in PbI$_2$, which suggests of the structural stability of MAPbI$_3$ML "on surface disposition". This observation reveals that the [PbI$_6$]$^{4-}$ octahedra in layered structures account for optical and electronic changes for hybrid perovskite systems, as discussed later, and can be one of the reasons for the experimentally observed features such as a shifted band edge emission. The structural differences caused by MoS$_2$ML on MAPbI$_3$ML deposition enhances the cohesive (binding) energies of the system as noted in Fig. 2.

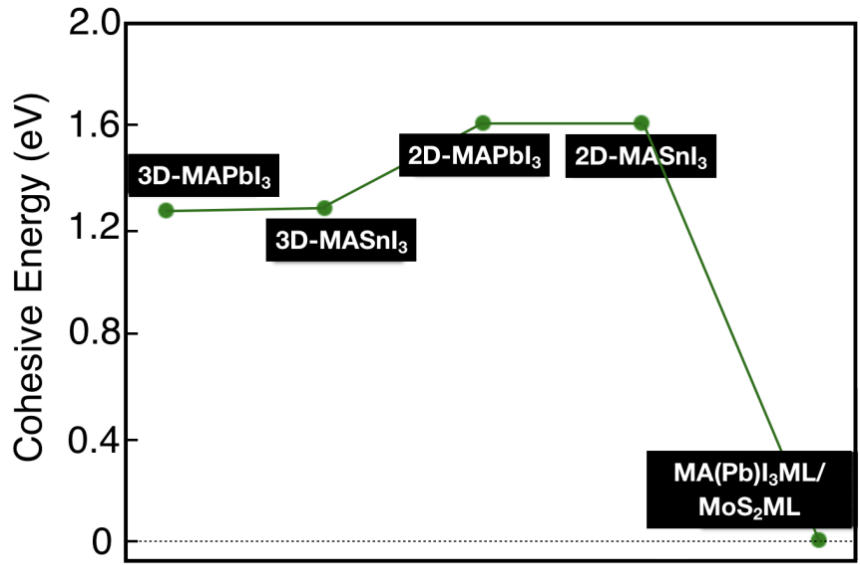

**Figure 2.** The relative ordering of cohesive energies (eV) of bulk and 2D MAPbI$_3$ and MAPbI$_3$ML/MoS$_2$ML (ML = monolayer).

*Electronic properties of bulk and 2D variants of MAYI$_3$ (Y=Pb, Sn):* Experimentally, the bandgap of cubic MAPbI$_3$ is ~ 1.5-1.62 eV [7]. Standard PBE calculations predict that bulk MAPbI$_3$ is a semiconductor with a direct bandgap of 1.73 eV, which agrees reasonably with the experiments. We perform additional band-structure calculation of 2D MAPbI$_3$ and the results indicate that both the 3D and 2D materials have a direct bandgap with the conduction band minimum (CBM) and valence band maximum (VBM) located at the M point of the Brillouin zone, as illustrated in Fig. 3. The energy states ranging from -3 to 3 eV are mostly contributed by Pb and I atoms, signifying their influence on the physical and chemical properties of these perovskites. Specifically, the states near the top of the valence band are dominated by I-*5p* and Pb-*6s* states, while the conduction bands are constituted by Pb-*6p* states with hybridization of I-*5p* states. As shown in Fig.S1, the surface terminated by PbI$_2$ contains more Pb atoms, which leads to a broader conduction band and thus a narrower bandgap of the 2D MAPbI$_3$ terminated by PbI$_2$.

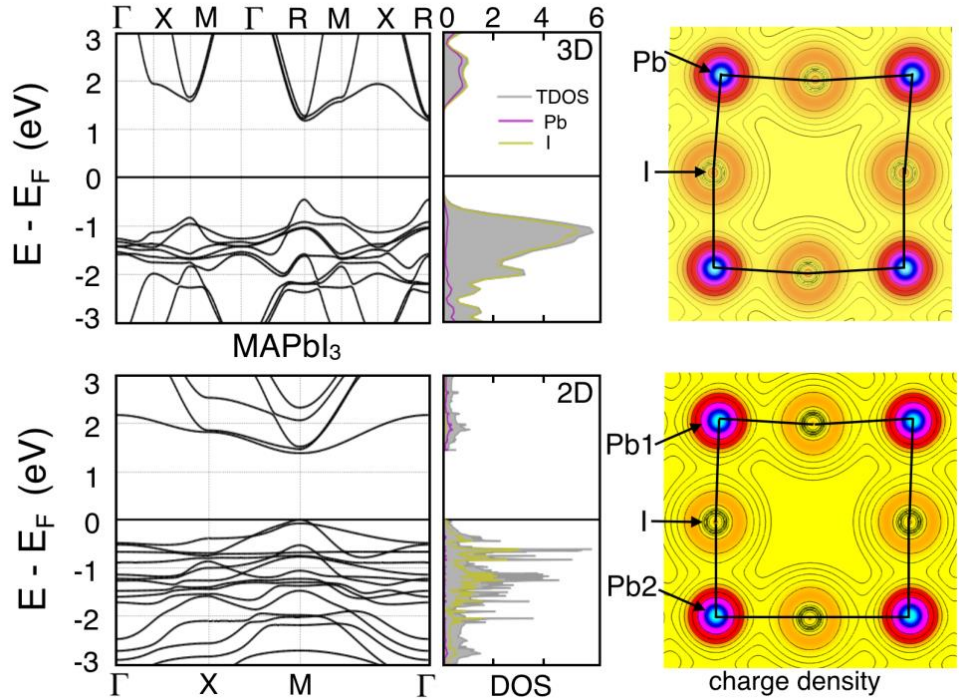

**Figure 3.** Comparative band-structure, density of states (DOS) and (001) projected charge density of bulk and [001] terminated 2D MAPbI$_3$. Transitioning from bulk to 2D contributes to the DOS being more structured and the bands flatter. The bulk and 2D MAPbI$_3$ possess a calculated direct bandgap of 1.73 eV and 1.27 eV at R-point and M-point of Brillouin zone, respectively. The [001] projected charge density of 2D MAPbI$_3$ reveals enhanced bonding between Pb and I (highlighted by non-circular lobes both at Pb and I sites) compared to bulk. Identical isosurface values are employed for the charge density plots.

For a deeper insight into the electronic properties of the heterostructured $MoS_2ML$, we compute the density of states (DOS), band-structure and (projected, full) charge densities for $MoS_2ML/MAPbI_3ML$, illustrated in Fig. 4a-f. The adsorption of $MAPbI_3ML$ introduces new flat energy levels between the valence and conduction bands of $MoS_2ML$ resulting in a bandgap of ~1.27 eV in Fig. 4a,b. As the interactions between $MAPbI_3ML$ and $MoS_2ML$ are weak, the band-structure of heterostructured $MoS_2ML$ is effectively a combination of those of $MoS_2ML$ and the adsorbed $MAPbI_3ML$. Hence, the bandgap reduction is attributed to the recombination of the energy levels from $MoS_2ML$ and $MAPbI_3ML$. Nevertheless, for the 2D assembly, the new energy levels appear in the region of the conduction band, indicating that $MoS_2ML$ can be tuned into a p-type semiconductor by doping with $MAPbI_3ML$. The DFT+PBE calculated charge density in Fig. 4c shows weak interaction between $MoS_2$ and $PbI_2$ layer.

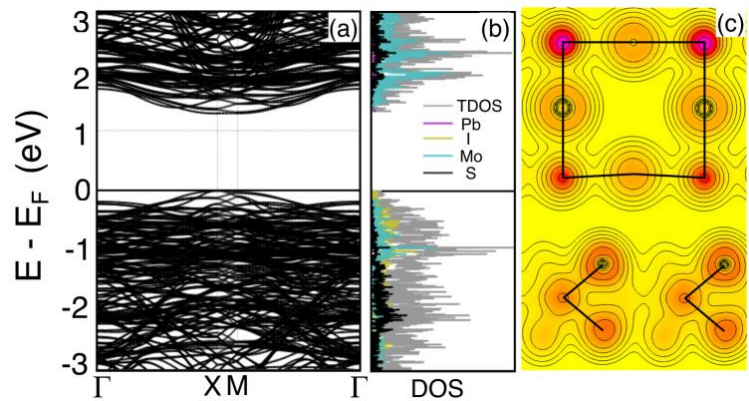

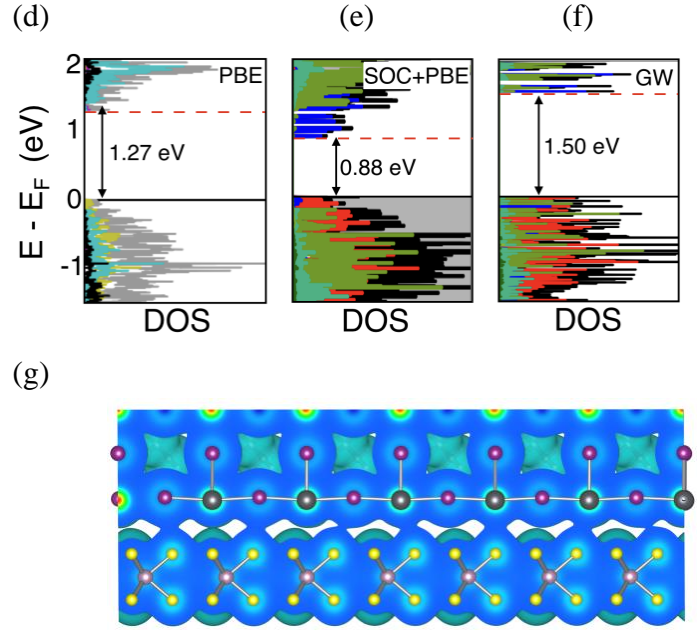

**Figure 4.** (a) The electronic band-structure, (b) density of states, and (c) (110) projected and full charge density of chemically deposited MAPbI$_3$ monolayer (ML) on MoS$_2$ML. Additionally, (d) the PBE, (e) SOC+PBE DOS are compared with (g) GW DOS predictions. The bandgap predictions from GW are slightly improved over PBE. (g) The full charge-density from GW calculations are also illustrated (bottom-layer is MoS$_2$ while the top-layer is PbI$_2$).

As shown in Fig. 4a,b for 2D heterostructure of MoS$_2$ML/MAPbI$_3$ML, the electronic states of MAPbI$_3$ML are localized near the Fermi-level in CBM and VBM, respectively. The MoS$_2$ states in band-structure predominantly lies below -0.2 eV with respect to the Fermi level, $E_F$. The conduction bands with an energy range from ~1 - 3 eV is derived largely from the Pb-$s/p$ and I-$s$ states. The valence bands with a range from -3 eV – 0 eV exert strong hybridization between the Pb-$p$, I-$p$, Mo-$d$ and S-$p$ states. The localized distribution of electron density on MAPbI$_3$ML and MoS$_2$ML indicates that quantum confinement due to reduced dimension from bulk to 2D heterostructures leads to optimal bandgap and stability of MoS$_2$ML/MAPbI$_3$ML. In Fig. 4d,e, we show density of states calculated using PBE and PBE+SOC (SOC=spin-orbit coupling). The reduced bandgap of ~0.9 eV in SOC+PBE compared to PBE case arises from the annihilation of the Pb-s and I-p states due to SOC. Whereas GW (SOC) bandgap in Fig. 4f for MAPbI$_3$ML/MoS$_2$ML is further improved to ~1.5 eV compared to PBE (1.27 eV) and PBE+SOC (~0.9 eV). The improved bandgap in GW calculation can be attributed to self-energy correction to Pb-s and I-p states. The charge density for MoS$_2$ML/MAPbI$_3$ML calculated using DFT-GW are shown in Fig.

4g. Notably, a weak charge sharing from MoS$_2$ to PbI$_2$ layer was observed in contrast to 2D projected charge-density in Fig. 4c. The effective charge pulled from S (top atoms)-layer to Pb-I layer in GW calculation make sense due to high electronegativity of I.

In Table 3, we compare predictions of bandgaps calculated with PBE and GW functionals, where the GW+SOC predicted bandgap remains within the optimal region 1.2 to 1.8 eV. Notably, the MoS$_2$ML (2D) has a predicted direct bandgap of 1.73 eV as shown in Fig. 3 (bottom-panel), closer to the experimentally observed bandgap of 1.70 eV [55,56]. The VBM of MoS$_2$ML is contributed by Mo-*4d* and S-*3p* states, while the CBM is mainly due to Mo-*4d* states and feebly by S-*3p* states.

**Table 3**. Bandgap (in eV) of 2D MAPbI$_3$ and MAPbI$_3$ML/MoS$_2$ML calculated with PBE, PBE+SOC, and GW (SOC), respectively.

| System | Bandgap [eV] | | |
|---|---|---|---|
| | **PBE** | **PBE + SOC** | **GW** |
| **2D MAPbI$_3$** | 1.50 | 0.41 | 1.60 |
| **MoS$_2$-ML/MAPbI$_3$** | 1.31 | 0.88 | 1.54 |

*Thermoelectric properties of bulk and 2D variants of MAYI$_3$ (Y=Pb, Sn):* A band-structure possessing a large-effective mass in CBM and with a minimum energy of about 5k$_B$T above the VBM, can potentially achieve nonmonotonic Seebeck coefficient, and hence a large thermoelectric power factor (*PF*) [57]. Deposition of MAPbI$_3$ML on MoS$_2$ML introduces resonant states that can achieve a tailored band-structure. We determine the band effective mass *m\** from the second order derivative of the band energy with respect to the wave vectors.

$$\left(\frac{1}{m^*}\right) = \frac{1}{\hbar^2} \frac{\partial^2 E_n}{\partial k_x \partial k_y}$$

where, *x* and *y* are the directions in reciprocal space, *n* is the band index, $E_n$ is the band energy, and $\hbar$ is the modified Planck's constant. The derivatives are evaluated at CBM for electrons ($m^e$) and at VBM for holes ($m^h$). The reduction in dimensions from bulk to 2D creates opportunities to optimize the highly anisotropic structures due to the smaller directional effective mass (electrical conductivity $\sigma \propto \frac{1}{m^*}$ ) [58].

The calculated effective hole/electron mass for bulk MAPbI₃ is (0.17; 0.19; 0.28) $m_h^*$ and (0.27; 1.51; 0.12) $m_e^*$, respectively. The three values in the parentheses represent the three perpendicular directions (x, y, z), while for 2D structures we consider the two planar directions (x, y). For 2D MAPbI₃, the corresponding values are (0.17; 0.33) $m_h^*$ and (1.53; 0.85) $m_e^*$, respectively. The calculated effective mass for 2D MAPbI₃ indicates an obvious in-plane anisotropy due to [100] stacking. The strong directional effective mass is larger along [100] stacking, while along the direction orthogonal to the stacking [010] the computations predict smaller values. Thus, we deduce that the carrier mobility is high (small effective mass) along [010], with heavy mass states present in the transverse direction, i.e., [100]. In MoS₂ML/MAPbI₃ML, the energy minima/maxima occur at the X [½ 0 0] point. The calculated hole- and electron-effective masses at M-point are $m_h^*$ = (0.76; 1.13) and $m_e^*$ = (3.40; 1.51), respectively. The large anisotropy in the effective mass along the [100] direction facilitates hole transport along [100] compared to the electrons. However, along the [010] direction, both electrons and holes have equal effective masses resulting in similar transport properties. As shown in Table 4, the effective mass in SOC case is slightly increased in X symmetry-direction, however, SOC shows slight reduced effective mass in M-symmetry point. Except M(x), the change in all other directions is negligible.

**Table 4**. Effective mass (in units of m₀) of MAPbI₃ML/MoS₂ML with and without spin-orbit coupling.

|  | NON-SOC | | | | SOC | | | |
|---|---|---|---|---|---|---|---|---|
|  | X | | M | | X | | M | |
|  | x | y | x | y | x | y | x | y |
| **Valence** | 0.759 | 1.127 | 3.838 | 0.183 | 0.902 | 1.702 | 1.666 | 0.981 |
| **Conduction** | 1.510 | 3.399 | 0.056 | 1.350 | 1.945 | 2.539 | 0.023 | 1.311 |

In thermoelectric (TE) materials, a high Seebeck coefficient (*S*) at a given carrier concentration results from a high overall DOS effective mass ($m_d$*). However, $\sigma$ decreases with increasing $m_d$*, and also depends on the inertial effective mass *m**. Ioffe showed empirically that a carrier concentration $n \sim 10^{18}$–$10^{20}$cm$^{-3}$ is mostly satisfied for doped semiconductors corresponding to degenerate semiconductors or semimetals [59]. Here, we predict a similar range for $n_{hole}$ and $n_{electron}$. As the doping concentration increases, $\sigma$ increases and $S$ decreases. Figs. 5 (a-c) present $S$ predictions for 2D MAPbI$_3$ and MoS$_2$ML/MAPbI$_3$ML. In comparison with bulk MAPbI$_3$ [60], we find that $S$ of 2D MAPbI$_3$ is ~20% lower. The large $S$ in bulk MAPbI$_3$ arises from the DOS and band mobility. As shown earlier in Fig. 3, the DOS above and below the Fermi level have a significant contrast in the bulk structures[53] compared to the 2D counterparts, and the smaller $m^*$ in 3D materials leads to an increased band mobility [60] that is important to enhance the electronic transport. In Fig. 5 (top panel), we show that bulk MAPbI$_3$ has marginally better TE properties than 2D MAPbI$_3$ and MoS$_2$ML/MAPbI$_3$ML as the contribution to the overall $S$ depends on the number of positive and negative charge carriers. At lower temperatures the population of minority carriers is small, and their contribution to $S$ is insignificant. Nevertheless, at higher temperatures, a broadening in Fermi distribution leads to an exponential increase in minority carrier conductivity, which reduces $S$.

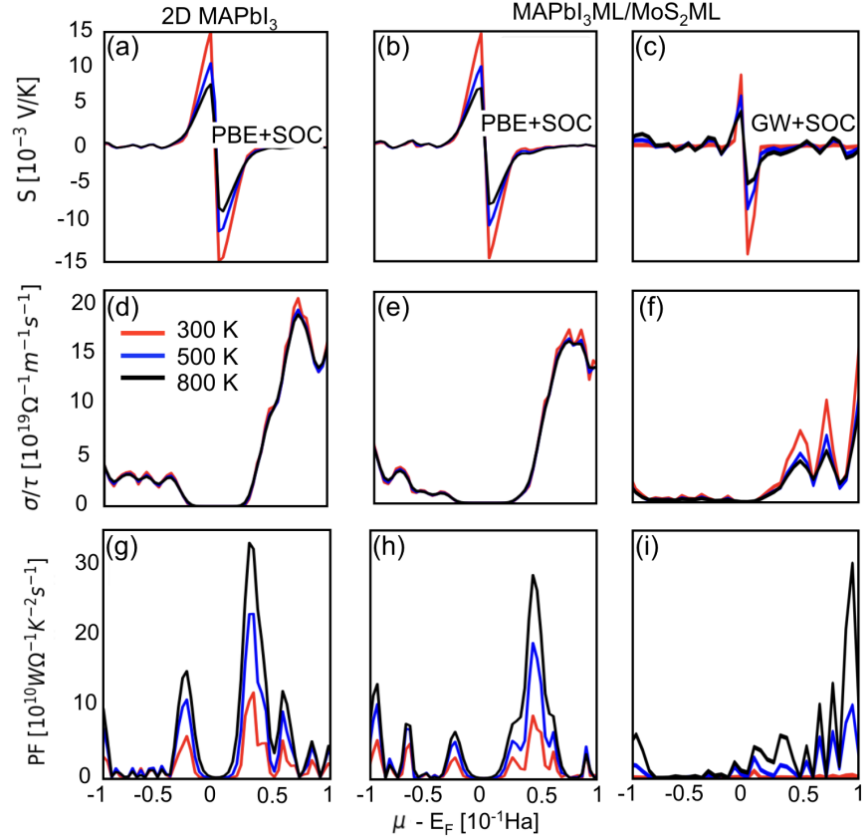

**Figure 5.** The Seebeck coefficient (a, b, c), the electrical conductivity (d, e, f), and the power factor (g, h, i) of 2D MAPbI$_3$ (left-panel) and MAPbI$_3$ML/MoS$_2$ML (center and right panels) are shown at 300 K, 500 K and 800 K, respectively (ML = monolayer). The thermoelectric properties of MAPbI$_3$ML/MoS$_2$ML calculated using SOC+PBE (b, e, h) and SOC+GW (c, f, i) are also presented as a function of the chemical potential.

$PF = S^2\sigma$ of a TE material quantifies their electrical power generation ability. The most efficient TE materials possess high $\sigma$ and high $S$. Since $\sigma$ and $S$ have competing dependencies on $n$, simultaneously obtaining high values for both properties is challenging. We compare the calculated $PF$ per relaxation time as a function of temperature for MAPbI$_3$ML and that with MoS$_2$ML. Although both materials originate from the same parent cubic crystal structure, differences are noted in their TE properties (Figs. 5 (d)–(i)). The dissimilarities in $\sigma$ of 2D MAPbI$_3$ and MoS$_2$ML/MAPbI$_3$ML originates from the intrinsic $sp$-type bonding of the heterostructure. The 2D OIHP deposited on MoS$_2$ has a lower $PF \sim 29.3 \times 10^{10}$ W/m.K$^2$.s relative to 2D MAPbI$_3$ ($\sim 32.5 \times 10^{10}$ W/m.K$^2$.s) due to an enhanced conductivity [60]. Both 2D MAPbI$_3$ and MoS$_2$ML/MAPbI$_3$ML exhibit strong temperature dependence for $S$, but $\sigma$ essentially remains temperature invariant. Thus, the *PF* predictions for specific chemical potential values at 800 K, Fig. 5(h) and (i), are ~100% and ~ 200% higher than at 500 K and 300 K, respectively, for both materials. Note that *PF* of 2D MAPbI$_3$ is ~10% higher than MoS$_2$ML/MAPbI$_3$ML, but the heterostructure has a significantly enhanced energetic stability. On one hand, OIHPs possess the advantage of hybridization, while on the other hand, their engineered geometries demonstrate the ability to tune TE properties through nanostructuring. In addition, the *PF* can be further controlled by the concentrations of the ionic dopants.

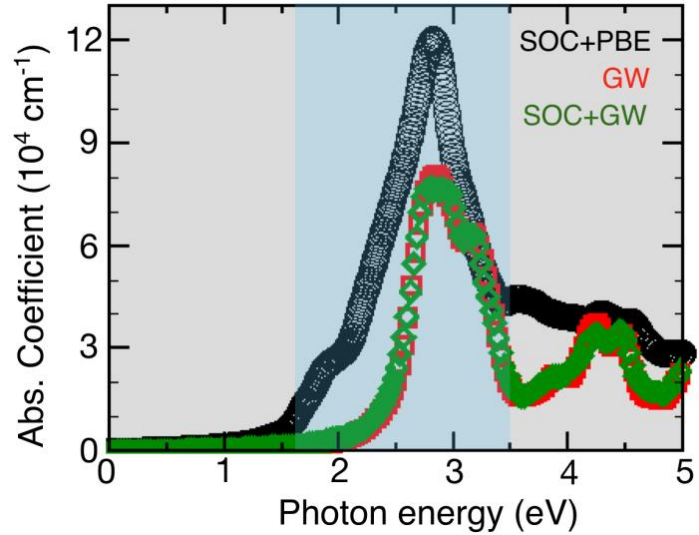

**Figure 6.** The calculated optical absorption spectra of the energetically most stable MAPbI$_3$ML/MoS$_2$ML. The heterostructured material exhibits very high absorption coefficient in the visible range of light from 1.59 to 3.26 eV (shaded zone) compared to Si and bulk MAPbI$_3$ [58-60] Predictions from SOC+PBE, GW and SOC+GW calculations are compared. The GW, a higher accuracy functional, shows a small decrease in the absorption coefficient, but the decrease is significantly less than an order of magnitude.

*Absorption spectra of MAPbI3/MoS2ML*: In Fig. 6, we display the computed optical absorption spectra for the energetically stable MoS$_2$ML/MAPbI$_3$ML heterostructure. The absorption spectra of the direct bandgap MoS$_2$ML/MAPbI$_3$ML is compared against that of two highly efficient solar-cell materials, Si and MAPbI$_3$ [53,54,61] Amongst these, the computationally designed MoS$_2$ML/MAPbI$_3$ML material exhibits a larger absorption coefficient in the visible-light range (from 1.59 to 3.26 eV) (for comparison, the optical absorption of Si and MAPbI$_3$ is elaborately discussed in Ref. [53,54,61]). Moreover, MoS$_2$ML/MAPbI$_3$ML shows modest absorption in infrared range because of its smaller bandgap relative to the bulk MAPbI$_3$, indicating that majority of the total solar irradiance spectrum can be absorbed by MoS$_2$ML/MAPbI$_3$ML. These desirable properties render the heterostructured MoS$_2$ML/MAPbI$_3$ML with the distorted perovskite chemistry as a very promising solar material with potential for high power conversion efficiency. The decomposition of MAPbI$_3$ when exposed to the ambient condition is a challenging issue, however, recent study by Fan and coworkers showed that MAPbI$_3$ thermally decomposes back to one crystal PbI$_2$ layer at a time driven by a surface dominated reaction [62]. The new understanding suggests that simple encapsulation with atomically thin 2D hexagonal boron nitride protects the top surface, which helps impeding the degradation of the underlying layers. Therefore, simply inhibiting the structural transitions of the surface layers using MoS2 capping by depositing on halide layers should greatly increase the overall structural and thermal stability.

**Conclusion**

In summary, we used density-functional theory methods to manipulate the electronic and thermoelectric properties of 2D organic-inorganic halide perovskites. Our study shows that a chemically deposited monolayer of 2D MAPbI$_3$ on a monolayer MoS$_2$ remarkably enhances the energetic stability and provides an optimal bandgap required for photovoltaic applications. The enhanced bandgap compared to bulk MAPbI$_3$ arises from the reduced dimension and increased quantum confinement effect in the 2D heterostructure. Our computations also reveal that MoS$_2$ML/MAPbI$_3$ML possesses excellent thermoelectric properties relative to bulk MAPbI$_3$ due to the intrinsic *sp*-type bonding of the heterostructure. The higher stability, suitable bandgap and better optical absorption promotes MoS$_2$ML/MAPbI$_3$ML as a potential material for the high-efficiency perovskite solar cells. We propose that 2D materials with charge alternations, such as transition metal dichalcogenides, offer a method to encapsulate the perovskite films. We believe our predictions provide a useful guideline for future experiments examining the stability and photovoltaic properties of 2D perovskites deposited on layered substrates.

**Acknowledgment**

The research was supported, in part, by the National Science Foundation (NSF) grant no. CMMI-1753770. The work at Ames Laboratory was supported by the U.S. Department of Energy (DOE), Office of Science, Basic Energy Sciences, Materials Science & Engineering Division, which is operated by Iowa State University for the U.S. DOE under contract DE-AC02-07CH11358.

**Author Contributions**

CRediT roles- RS: Calculations, Formal analysis; Writing - original draft; Writing - review & editing. PS: Conceptualization, Formal analysis; Writing - original draft; Writing - review & editing. GB: Funding acquisition; Supervision; Resources; Writing - review & editing.

**Data availability**

The authors declare that the data supporting the findings of this study are available within the paper and supplement. Also, the data that support the plots within this paper and other finding of this study are available from the corresponding author upon reasonable request.

# Supplementary information

# Effect of heterostructure engineering on electronic structure and transport properties of two-dimensional halide perovskites

Rahul Singh,[1] Prashant Singh[2] and Ganesh Balasubramanian[3]

[1]Department of Mechanical Engineering, Iowa State University, Ames, IA 50011
[2]Ames Laboratory, U.S. Department of Energy, Iowa State University, Ames, IA 50011
[3]Department of Mechanical Engineering & Mechanics, Lehigh University, Bethlehem, PA 18015

Below we provide the supporting material referenced in the main article, as well as other results that support the research methodologies and the conclusions.

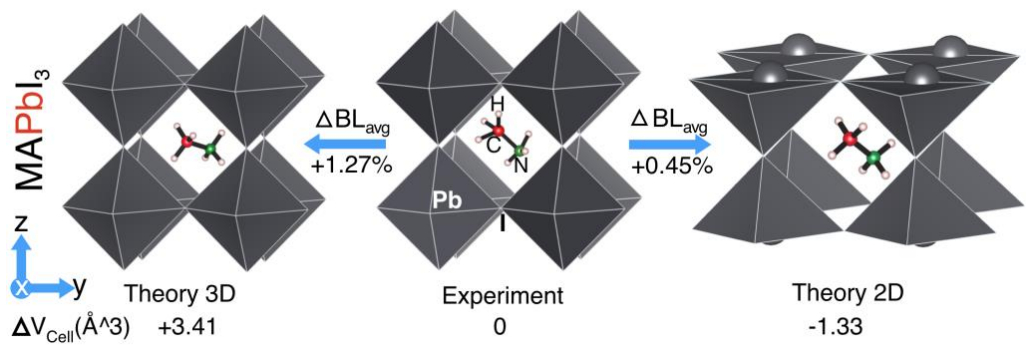

**Figure S1.** The side view of bulk MAPbI$_3$ unit cell as obtained from (a) simulations and (b) experiments. The side view of (c) [001] terminated 2D MAPbI$_3$ unit cell. The unit cell volume of the bulk and energetically relaxed structure increases by 3.41% relative to the experimentally measured volume, while the 2D cell shrinks by 1.33%. In comparison to the experiments the average Pb-I bond-length of bulk and 2D MAPbI$_3$ increases by 1.27 and 0.45%, respectively. For the relaxed 2D MASnI$_3$, the unit cell volume and average Sn-I bond length increases by 1.65 and 2.05%, respectively, compared to bulk.

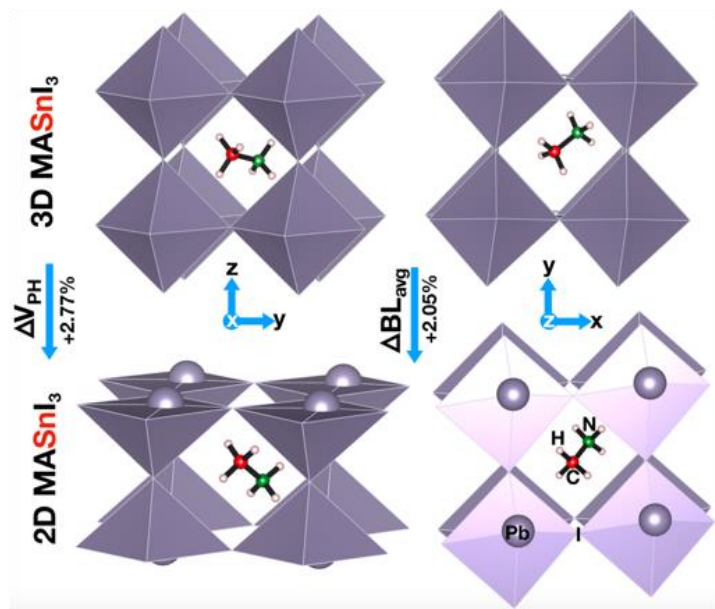

**Figure S2.** The side view of bulk MASnI$_3$ unit cell as obtained from simulations and the side view of [001]) terminated 2D MASnI$_3$ unit cell. The volume of relaxed of 2D MASnI$_3$ increases by 2.77% as compared to bulk MASnI$_3$. Similarly, Sn-I bond length increases by 2.05% compared to bulk.

Theoretical predictions suggested that the thermoelectric efficiency could be greatly enhanced by quantum confinement through ionic doping or heterostructure engineering. The heterovalent or isovalent metal ions such as ($Sn^{2+}$, $Cd^{2+}$, $Zn^{2+}$, $Mn2+$) have been introduced as dopants with the possibility of imparting paramagnetism to increase stability.

*Structural optimization*: For Mn-doped 2D $MAPbI_3$ shown in Fig. S3, the calculated (experimental[48,49]) lattice parameters are $a = 6.437$ (6.361) Å and $b = 6.449$ (6.361) Å, respectively. The doped structure shows 25% distortion in Mn-I polyhedra, while only 9% distortion in Pb-I polyhedra. The average Mn-I bond-length increases by 4% relative to Pb-I.

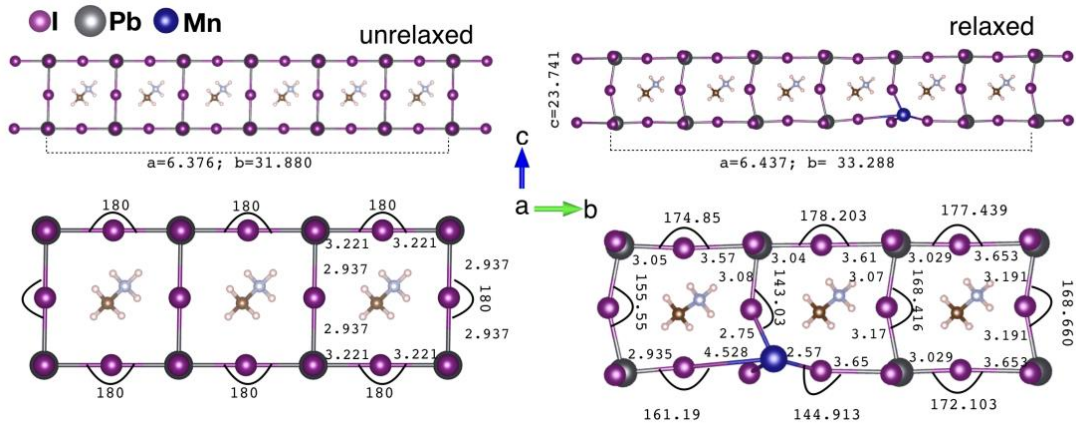

**Figure S3.** The unrelaxed (left-panel) and optimized structure (right-panel) of Mn-doped $MAPbI_3ML$.

*Electronic structure Mn-doped 2D MAPbI₃*: The spin-polarized band structure for 10% Mn-doped 2D MAPbI$_3$ are presented in Fig. S4. The bandgap of the Mn-doped OIHP at Pb-site is 1.2 eV, while Mn at Sn site reduces the bandgap to ~1 eV. Below, we discuss only the effect of Mn-doping at Pb-site.

Both the top of the valence band and the bottom of the conduction band are located at the M [½ ½ 0] point. The doping results in a more stable structure by distorting the neighboring environment, see charge density plot in Fig. S4 (b) (top). However, the bandgap of Mn-doped OIHP decreases due to large structural distortion. The Mn spin-up channel is completely filled, and down-spin channel is completely empty leading to strong magnetic behavior (Mn-moment ~ 5 $\mu_B$) as shown by magnetization density plot in Fig. S4 (b) (bottom). The top flat band just below the Fermi level, corresponds to the Mn impurity band, which is attributed to the Mn-*3d* state. The Mn-doping changes the electronic structure near the Fermi level affecting stability as well as the transport properties. While there is no notable improvement in transport properties, the stability of the structure is enhanced by Mn-doping.

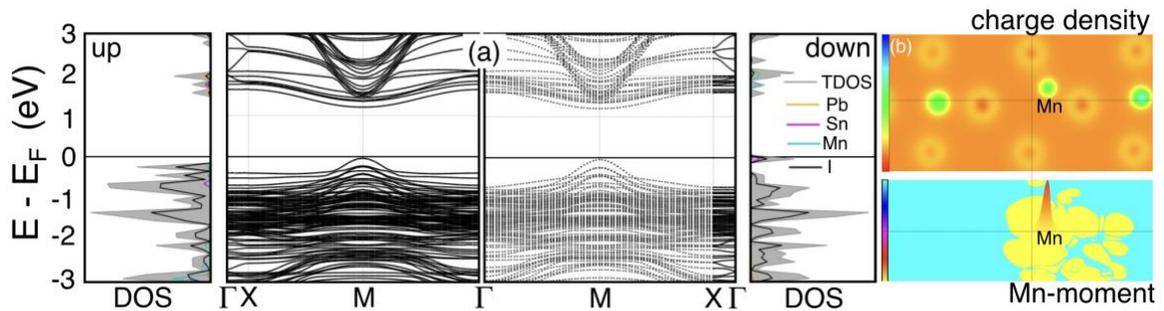

**Figure S4.** (Color online) (a) Spin-polarized band-structure, (b) charge density (top) and magnetization density (bottom) of Mn-doped (at Pb-site) 2D MAPbI$_3$ monolayer stacked along [010] direction. In Mn-doped MA(Pb;Sn)I$_3$, the strong magnetic characteristic of Mn with high saturation magnetization of ~5 Bohr magneton ($\mu_B$), leads to induced *I* moments on nearby sites by distortion of the MnI$_2$ polyhedra. Mn-doping stabilizes MAPbI$_3$ML almost by ~0.12 eV with respect to 2D MAPbI$_3$ and 2D MASnI$_3$.

*Effective mass and thermoelectric properties of Mn-doped 2D MAPbI₃*: The Mn-doped 2D MAPbI3 shows large anisotropy in effective mass along [100] stacking at Pb-site [(0.12; 5.42) $m_h^*$, (2.50; 1.15) $m_e^*$]. The large anisotropy in the effective mass along the [100] direction facilitates hole transport holes along [100]. However, along the [010] direction, both electrons and holes experience equal effective mass resulting in similar transport properties. Regardless of the enhanced stability, the transport properties (see in Figure S5) are poor.

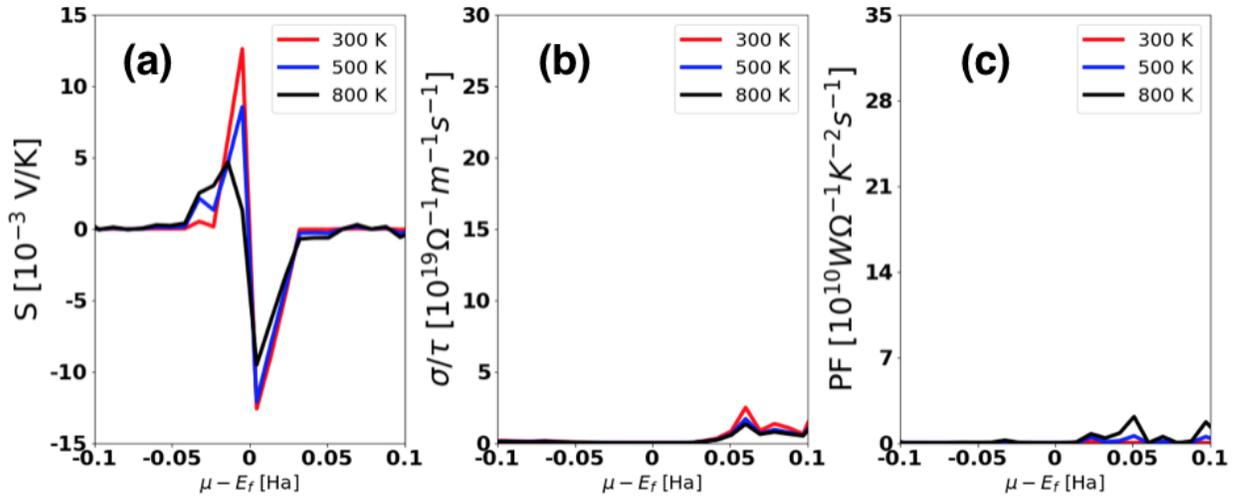

**Figure S5.** The (a) Seebeck coefficient, (b) electrical conductivity and (c) and power factor for 10% Mn-doped 2D MAPbI3. The Seebeck coefficient is lower relative to the predictions for 2D MAPbI3 and MAPbI3ML/MoS2ML. However, there is a factor of 10 drop in electrical conductivity and power factor compared to corresponding values for 2D MAPbI3 and MAPbI3ML/MoS2ML.